\documentclass[a4paper]{jpconf}

\usepackage{graphicx}% Include figure files
\usepackage{dcolumn}% Align table columns on decimal point
\usepackage{bm}% bold math
\usepackage{amsmath}	% Advanced maths commands
\usepackage{amssymb}	% Extra maths symbols

\begin{document}
\title{The population of highly magnetized neutron stars}

\author{R.O. Gomes, }
\address{Department of Astronomy, Universidade Federal do Rio Grande do Sul, Porto Alegre, Brazil}

\author{V. Dexheimer }
\address{Department of Physics, Kent State University, Kent OH 44242 USA}

\author{B. Franzon}
\address{FIAS, Ruth-Moufang 1, 60438 Frankfurt am Main, Germany}

\author{S. Schramm}
\address{FIAS, Ruth-Moufang 1, 60438 Frankfurt am Main, Germany}

\ead{rosana.gomes@ufrgs.br}

\begin{abstract}
In this work, we study the effects of strong magnetic field configurations on the population of neutron stars. 
The stellar matter is described within a relativistic mean field formalism 
which considers many-body force contributions in the scalar couplings.  
We choose the parametrization of the model that reproduces nuclear matter properties at saturation and 
also describes massive hyperon stars. 
Hadronic matter is modeled at zero temperature, in beta-equilibrium, charge neutral and populated by the baryonic octet, 
electrons and muons. 
Magnetic effects are taken into account in the structure of stars by the solution of the Einstein-Maxwell equations with the 
assumption of a poloidal magnetic field distribution. 
Our results show that magnetic neutron stars are populated essencialy by nucleons and leptons, due to the fact that strong magnetic fields 
decrease the central density of stars and, hence, supress the appearance of exotic particles.

\end{abstract}

\section{Introduction}
\hfill \break

The topic of strong magnetic fields in neutron stars have been extensively studied in the literature. Magnetic field effects have been 
explored both on the equation of state \cite{Chakrabarty:1997ef,Broderick:2000pe,Broderick:2001qw,Sinha:2010fm,Lopes:2012nf,
Casali:2013jka,Gomes:2014dka,Gao:2015jha,PerezMartinez:2007kw,Orsaria:2010xx,Dexheimer:2012mk,Denke:2013gha,Isayev:2015rda,
Felipe:2010vr,Paulucci:2010uj,Rabhi:2009ih,Dexheimer:2012qk} and on the strucuture of magnetic neutron stars 
\cite{Bonazzola:1993zz,Bocquet:1995je,Cardall:2000bs,Chatterjee:2014qsa,Franzon:2015sya}, in order to identify their impact 
on the global properties of neutron stars, such as masses and radii.
In particular, Refs. \cite{Chatterjee:2014qsa,Franzon:2015sya} have shown that magnetic effects in the equation of state are not 
strongly significant for the determination of global properties of stars (i.e., mass and radius), although magnetic field dependence
on microscopic processess as neutrino emission and consequently the cooling are still important \cite{Dexheimer:2012}.

\newpage
Another relevant effect of strong magnetic fields in neutron stars is their equatorial radius enhancement and the consequent decrease of 
the central baryon densities \cite{Franzon:2015sya}. Such effect can have a severe impact on the population of highly 
magnetized neutron stars. The decrease of the central chemical potential, 
related to the baryon density, can prevent the appearance of phase transitions \cite{Franzon:2015sya,Franzon:2016} as well as 
the existence of exotic degrees of freedom. In this work, we show for a particular parametrization of the many-body forces model (MBF model) that 
the hyperon population of neutron stars is completely supressed in the presence of central magnetic fields $B_c \sim 10^{18}\,\mathrm{G}$.

\section{Hadronic Model}
\hfill \break

In order to describe hadronic matter inside magnetic stars, we make use of a many-body forces model (MBF model) \cite{Gomes:2014aka}, 
which is a relativistic mean field model that takes into account nonlinear meson-meson contribuitions in the scalar interactions. 
The lagrangian density of the MBF model reads \cite{Gomes:2014aka}:
\begin{equation}\begin{split}\label{lagrangian}
\mathcal{L}&= \underset{b}{\sum}\overline{\psi}_{b}\left[\gamma_{\mu}\left(i\partial^{\mu} -g_{\omega b}\omega^{\mu} -g_{\phi b}\phi^{\mu} 
-g_{\varrho b}\mathbf{\textrm{\ensuremath{I_{3b}}\ensuremath{\varrho_3^{\mu}}}}\right)
-m^*_{b \zeta}\right]\psi_{b} +\underset{l}{\sum}
\overline{\psi}_{l}\gamma_{\mu}\left(i\partial^{\mu}
-m_{l}\right)\psi_{l} 
\\& +\left(\frac{1}{2}\partial_{\mu}\sigma\partial^{\mu}\sigma-m_{\sigma}^{2}\sigma^{2}\right)
+\frac{1}{2}\left(-\frac{1}{2}\omega_{\mu\nu}\omega^{\mu\nu}+m_{\omega}^{2}\omega_{\mu}\omega^{\mu}\right)
+\frac{1}{2}\left(-\frac{1}{2}\phi_{\mu\nu}\phi^{\mu\nu}+m_{\phi}^{2}\phi_{\mu}\phi^{\mu}\right)
\\& +\frac{1}{2}\left(-\frac{1}{2}\boldsymbol{\varrho_{\mu\nu}.\varrho^{\mu\nu}}+m_{\varrho}^{2}\boldsymbol{\varrho_{\mu}.\varrho^{\mu}}\right)
+\left(\frac{1}{2}\partial_{\mu}\boldsymbol{\delta.}\partial^{\mu}\boldsymbol{\delta}-m_{\delta}^{2}\boldsymbol{\delta}^{2}\right),
\end{split}\end{equation}
where the indices $b$ and $l$ denote the degrees of freedom of baryons ($p^+$, $n$, $\Lambda$, $\Sigma^+$, $\Sigma^0$, $\Sigma^-$, $\Xi^0$, $\Xi^-$) 
and leptons ($e^-$, $\mu^-$), respectively.
The first line represents the Dirac lagrangian for baryons and leptons, respectively. 
The second line presents the lagrangian densities of the scalar-isoscalar $\sigma$ field and the vector-isoscalar $\omega$ and $\phi$ fields, 
which reproduce the attractive and repulsive features of nuclear interaction. 
The isovector fields $\delta$ (scalar) and $\varrho$ (vector) are introduced in the third line, 
and are responsible for the description of isospin asymmetry present in neutron stars. 
The inclusion of the $\phi$ allows for a more accurate description of the hyperon-hyperon interaction. In this work we do not 
include the $\sigma^*$ meson in order to focus on massive hyperon stars (for more results of the MBF model and the inclusion of the 
$\sigma^*$ meson, see Ref. \cite{Gomes:2014aka}).

The many-body forces contributions are introduced in the effective couplings of the scalar mesons: 
\begin{equation}
g^{*}_{\sigma b} =\left(1+ \frac{g_{\sigma b}\sigma+ g_{\delta b}I_{3b}\delta_{3}}{\zeta m_{b}}  \right)^{-\zeta} g_{\sigma b},
\qquad
g^{*}_{\delta b} =\left(1+ \frac{g_{\sigma b}\sigma+ g_{\delta b}I_{3b}\delta_{3}}{\zeta m_{b}}  \right)^{-\zeta} g_{\delta b},
\label{geff}
\end{equation}
making them no longer constants, but dependent on the scalar fields. 
Consequently, the effective baryon mass is also affected by many-body contribuitions as: 
$m^*_{\zeta b} = m_b - g^*_{\sigma b}\sigma - g^*_{\delta b}I_{3b}\delta_{3}$.
The parameter $\zeta$ regulates the intensity of the nonlinear contributions from the scalar fields, which are interpreted as 
higher order meson-meson interactions contributions. 

In order to describe massive hyperon stars, we use the parametrization $\zeta=0.040$ that reproduces a 
binding energy per nucleon $B/A = -15.75\,\mathrm{MeV}$ and a saturation density $\rho_0 = 0.15\,\mathrm{fm^{-3}}$.  
Nuclear matter properties values at saturation for this choice of parameters are: 
effective mass of the nucleon $m^*_n = 0.66m_n$, compressibility $K_0=297$ (MeV),  
symmetry energy $J_0=32\,\mathrm{MeV}$ and slope of the symmetry energy $L_0=97\,\mathrm{MeV}$ \cite{Gomes:2014aka}. 
Also, we use the spin-flavor SU(6) symmetry in order to define the hyperon couplings with the 
$\omega$, $\phi$, $\varrho$ and $\delta$ fields \cite{hyperon1,hyperon2}, and fix the hyperon potentials depths 
$U_{\Lambda}^{N}=-28\,\mathrm{MeV}$, $U_{\Sigma}^{N}=+30\,\mathrm{MeV}$ and $U_{\Xi}^{N}=-18\,\mathrm{MeV}$ 
to obtain their coupling with the $\sigma$ field.

\section{Structure of magnetic stars}
\hfill \break

In order to describe the structure of magnetic neutron stars, we solve the Einstein-Maxwell equations system.  
In particular, we use the open source LORENE C++ library for numerical relativity developed by 
Bonazzola et al. \cite{Bonazzola:1993zz,Bocquet:1995je}.

This formalism solves the Einstein's equations coupled to the Maxwell's equations 
for an axisymmetric metric using the $3 + 1$ decomposition, which is a 
typical tool for numerical relativity.
The system of equations is solved under the assumptions of meridional currents absence, 
infinite conductivity, magnetostatic equilibrium (from momentum and energy conservation) 
and for a poloidal magnetic field distribution. 
As the electromagnetic tensor fulfills the homogeneous Maxwell equations, 
only the Maxwell-Gauss and Maxwell-Amp\`ere equations are left to be solved. 
For more details on the formalism, 
see Refs. \cite{Bonazzola:1993zz,Bocquet:1995je,Chatterjee:2014qsa,Franzon:2015sya}.

The formalism used in this work does not include the dynamics that would 
ultimately give rise to a magnetic field in the stars.   
For this reason, the magnetic field is generated by a current, 
which is obtained from the EoS together with a current function $j_0$, which is introduced as a parameter. 
It has been shown in previous works that magnetic effects on the equation of state do not 
play a significant role on the determination of macroscopic properties of highly 
magnetized neutron stars \cite{Chatterjee:2014qsa,Franzon:2015sya} and in this work, for this reason, 
we do not include such effects on the equation of state of the MBF model.

\section{Maximum mass stars population}
\hfill \break

Strong magnetic fields have a very large impact on the population of neutron stars. 
First, the Lorentz force in magnetic stars acts agaist gravity, increasing the radius of stars and, 
consequently, reducing substantially their central density (and respective central baryon chemical potential).
This can be seen in Fig.~1, in which the central density of a $M_B=2.2$ M$_\odot$ baryon mass star is shown 
as function of different choices of the current function $j_0$, i.e. for different magnetic field distributions.

\begin{figure}[t!]
\centering
\includegraphics[width=15cm, height=9cm]{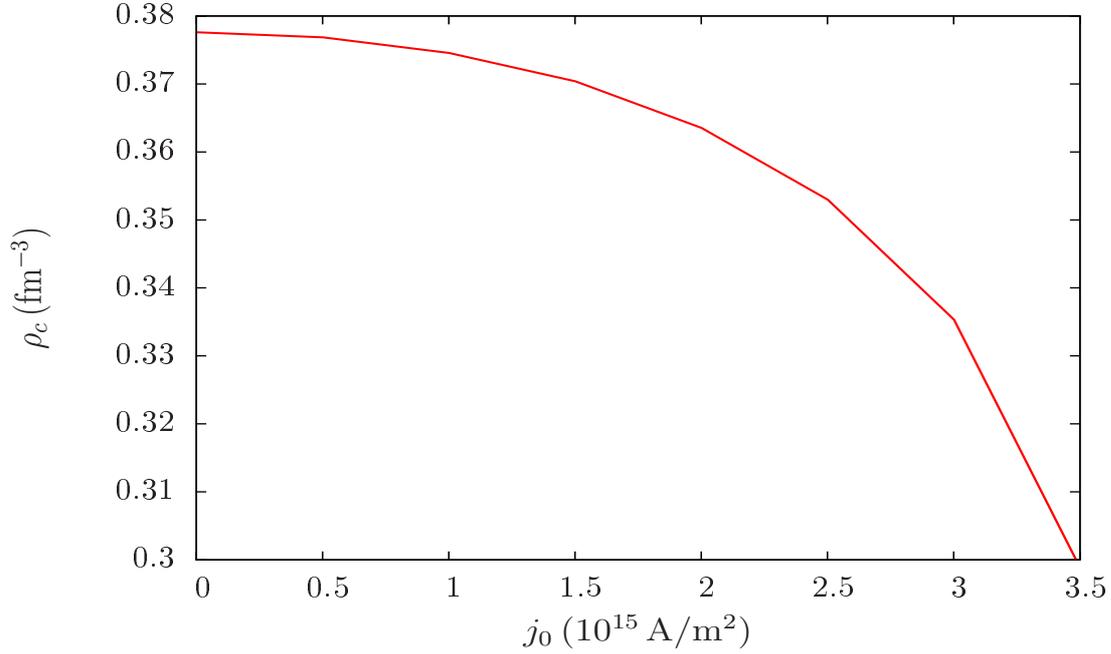}
\hspace{2pc}
\caption{\label{label} Central baryon density (vertical axis) of a $M_B=2.2$ M$_\odot$ baryon mass star as a function of the current function (horizontal axis). }
\end{figure}

We can now turn our attention to the maximum gravitational mass stars generated with our EoS including or not magnetic fields. 
Fig.~2  shows the particle population as a function of the radius for the non magnetic star, which is spherical and 
has a gravitational mass $M_G=2.15$ M$_\odot$ and central density $\rho_c = 0.86\,\mathrm{fm^{-3}}$. 
The population is dominated by $n$, $\Lambda$ and $\Xi^-$ particles in the inner region, due to the high densities 
reached at the stellar core \cite{Gomes:2014aka}.

Fig.~3 shows the population for the margnetic star that has a gravitational mass $M_G=2.22$ M$_\odot$ and  
the central density $\rho_c = 0.70\,\mathrm{fm^{-3}}$. 
This is the maximum gravitational mass star for a magnetic configuration given by a current function 
$j_0 = 3.5 \times 10^{15}\,\mathrm{A/m^2}$, which describes the strongest magnetic field distribution 
allowed by the code for the MBF model, with surface $B_s = 3.8 \times 10^{17}\,\mathrm{G}$ and 
central $B_c = 1.1 \times 10^{18}\,\mathrm{G}$ magnetic fields values. 
The particle population for the magnetic case is shown as a function of the polar radius of the star, 
but the results are qualitatively the same for the equatorial direction. 
It can be seen that, in case of central magnetic fields of the order $\sim 10^{18}\,\mathrm{G}$, the hyperon  
population is completely supressed. 

The decrease of strangeness at the core of magnetic stars is directly caused by the 
decrease of their central density. In such stars, the central region 
does not provide enough energy to reach the hyperons threshold of appearance. 
Similar results have already been discussed in the literature for highly magnetized hybrid stars, 
proving that strong magnetic fields can also prevent the occurance of phase transitions 
in such objects \cite{Franzon:2015sya,Franzon:2016}. 
Furthermore, although not investigated in this work, the microscopic effects of  
Landau quantization in the EoS also shifts the threshold of the appearance of hyperons to higher 
densities due to the extra magnetic contribuition to the particles energy 
levels \cite{Broderick:2000pe,Gomes:2014dka,Strickland:2012vu}.

\begin{figure}
\begin{minipage}{18pc}
%\begin{center}
\centering
\includegraphics[width=20pc, height=6cm]{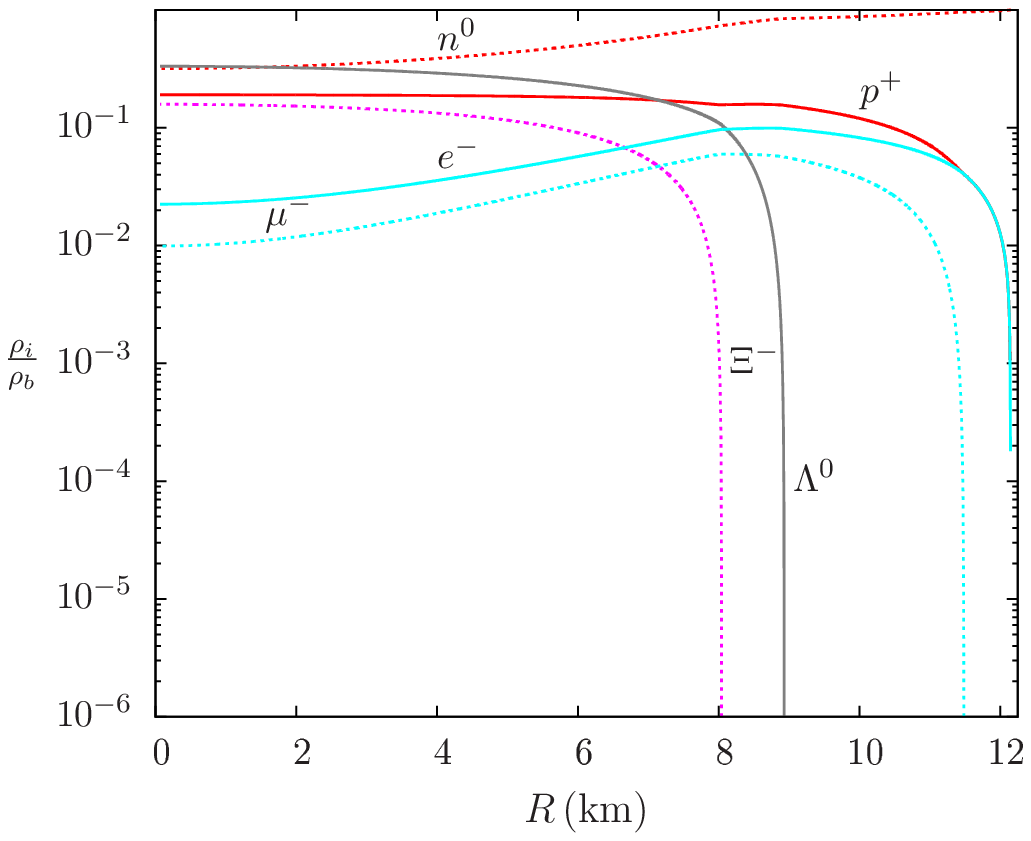}
\caption{\label{label} Particle stellar population for the maximum achieved gravitational mass star ($M_G=2.15$ M$_\odot$) in the 
non magnetic case. The vertical axis shows the particles densities normalized by the baryon density and the 
horizontal axis shows the radius of the star.}
\end{minipage}
%\end{center}
\hspace{2pc}%
\begin{minipage}{18pc}
\includegraphics[width=20pc, height=6cm]{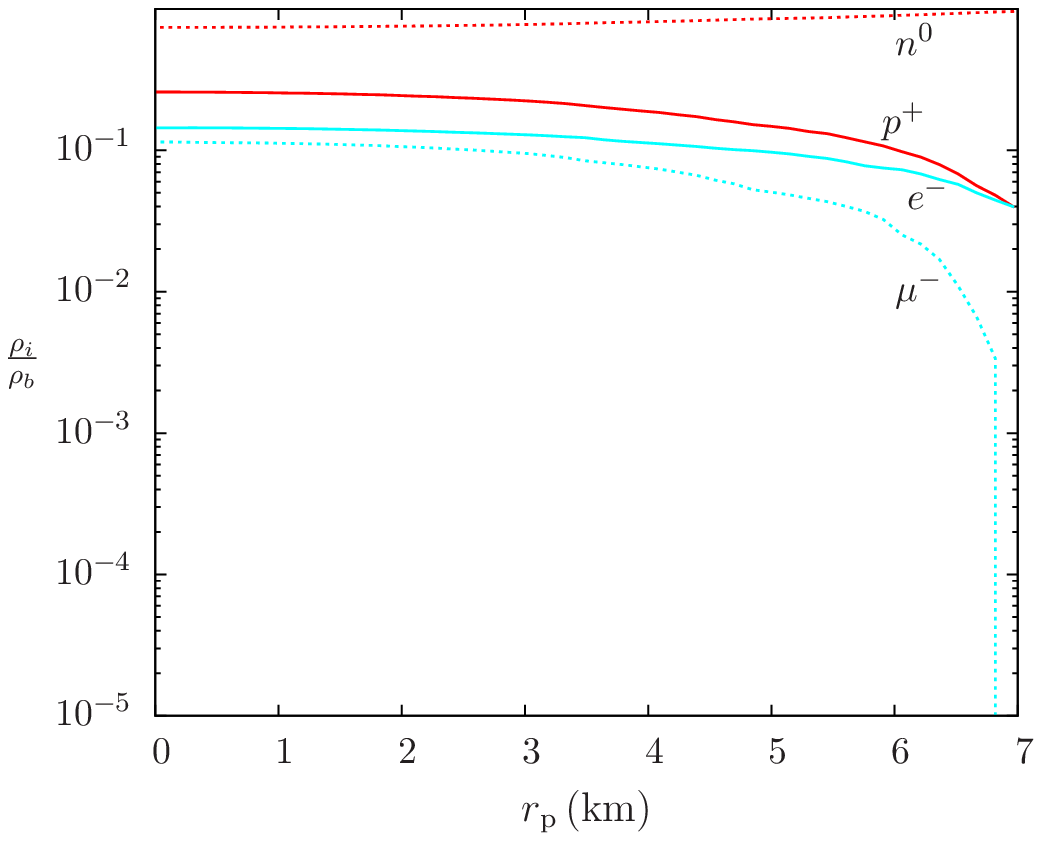}
\caption{\label{label} Particle stellar population for the maximum achieved gravitational mass star ($M_G=2.22$ M$_\odot$) in the 
magnetic case ($j_0 = 3.5 \times 10^{15}\,\mathrm{A/m^2}$). The vertical axis shows the particles densities normalized by the baryon density and the 
horizontal axis shows the polar radius of the star.}
\end{minipage} 
\end{figure}

\newpage

\section{Discussion}
\hfill \break

Strong magnetic fields in neutron stars have long been studied, usually focusing on 
their effects on the global properties of stars, such as masses, radii and deformation. 
In this work, we have applied a formalism that allows for describing consistently the structure of 
magnetic neutron stars, taking into account their deformation as well as solving the 
Einstein-Maxwell equations system for determining their strutucture 
(instead of using the Tolman-Oppenheimer-Volkoff spherical solutions).
The hadronic matter interaction inside the stars was described by the MBF model, which considers many-body 
forces contributions in the effective coupling of scalar mesons. The matter was modeled to be at zero temperature, 
charge neutral, in beta-equilibrium and populated by nucleons, hyperons and leptons (for the non magnetic case). 

We have shown that in the presence of strong magnetic field distributions, the central density of stars 
decreases due to the Lorentz force. This has the direct impact of decreasing the strangeness fraction at their core. 
Comparing the case of the maximum gravitational mass for the scenarios of a non magnetic star and for the highest magnetic 
field configuration allowed by the code,  we find that central magnetic fields as strong as $\sim 10^{18}\, \mathrm{G}$ 
supress completely the hyperon population of magnetic neutron stars, similarly to previous results for 
phase transitions in hybrid stars. 
These results imply that, as magnetic field strengths decay over time, a repopulation 
mechanism is driven inside stars, similarly to the one investigated for slowing down pulsars \cite{Negreiros:2011ak}.
Such results are important for the detection of signals for exotic matter (hyperons, delta isobars, quark matter)  
inside magnetic neutron stars such as their cooling, tidal deformation and gravitational wave emisions.
A similar work considering  other parametrizations of the MBF model and performing a similar population investigation 
but for atars with the same baryonic mass is already in preparation.

\end{document}